\documentclass[
 reprint,
 amsmath,amssymb,
]{revtex4-2}

\usepackage{graphicx}
\usepackage{dcolumn}
\usepackage[dvipsnames]{xcolor}
\usepackage{bm}
\usepackage{hyperref}
\usepackage[T1]{fontenc}
\usepackage{amsmath}

\hypersetup{
    colorlinks=true,
    linkcolor=blue,
    filecolor=magenta,      
    urlcolor=cyan,
    pdftitle={Overleaf Example},
    pdfpagemode=FullScreen,
    }
    
\begin{document}

\preprint{APS/123-QED}

\title{Shaping potential landscape for organic polariton condensates in double-dye cavities} 

\author{Anton D. Putintsev$^{1}$, Kirsty E. McGhee$^{2}$, Denis Sannikov$^{1}$, %\thanks{Correspondence address: Anton.Putintsev@skoltech.ru}
%\author{
Anton V. Zasedatelev$^{1}$, Julian D. T\"{o}pfer$^{1}$, Till Jessewitsch$^{3}$, Ullrich Scherf$^{3}$, David G. Lidzey$^{2}$, and Pavlos G. Lagoudakis$^{1}$ }

\affiliation{$^1$Hybrid Photonics Laboratory, Skolkovo Institute of Science and Technology, Territory of Innovation Center Skolkovo, Bolshoy Boulevard 30, building 1, 121205 Moscow, Russia}
\affiliation{$^2$Department of Physics and Astronomy, University of Sheffield, Hicks Building, Hounsfield Road, Sheffield S3 7RH, UK}
\affiliation{$^3$Macromolecular Chemistry Group and Institute for Polymer Technology, Bergische Universität Wuppertal, Wuppertal, Germany}
%\affiliation{$^{3}$Department of Physics and Astronomy, University of Southampton, Southampton, UK}

\date{\today}

\begin{abstract}

We investigate active spatial control of polariton condensates independently of the polariton-, gain-inducing excitation profile. This is achieved by introducing an extra intracavity semiconductor layer, non-resonant to the cavity mode. Saturation of the optical absorption in the uncoupled layer enables the ultra-fast modulation of the effective refractive index and, through excited-state absorption, the polariton dissipation. Utilising these mechanisms, we demonstrate control over the spatial profile and density of a polariton condensate at room temperature.

\end{abstract}
                              
\maketitle

The ability to shape the potential energy landscape to confine polariton condensates into geometrically arranged regular or arbitrary arrays underpins various applications such as topological polaritonics \cite{Dang22,Ardizzone2022,StJean2021,Kravtsov2020}, lattice simulators \cite{Kavokin2022,Berloff2017,AMO2016,Carusotto2013}, neural networks \cite{Ballarini2020, Mirek2021}, and allows for the investigation of fundamental physics \cite{Gnusov2023, Topfer2020}. An essential part in all these studies is to have tools and techniques to control the polariton energy landscape that is usually achieved by means of lithographic or optical methods. The lithographic approach allows for the local control of the photonic component of the system by a straightforward lithographic modulation of the cavity length \cite{elDaif2006}, deep etching of micropillars into the structure \cite{Obert2004,Bajoni2008}, metal films deposition on the surface of a grown microcavity structure \cite{Kim2008}, or by exploiting natural photonic disorder potentials present in microcavities due to growth related defects \cite{Kulakovskii2010,Zajac2012}. 
All-optical methods to control the energy landscape of polaritons offer higher flexibility in customizing its shape whilst circumventing inherent disorder. Whether it is a single gaussian pump or an array of pumps geometrically arranged in multiple-spot configurations and lattices,  forming either ballistically expanding \cite{Alyatkin2020} or trapped polariton condensates \cite{Wei2022}, one exploits a localized blueshift effect which at the same time induces additional gain in the system.

Manipulating polariton potentials through additional gain inducing optical excitation adds a real-valued component to the interaction between polariton condensates that strongly affects the energy of the condensate. An alternative approach of utilising local control of polaritons through dissipation via ion implantation was recently proposed \cite{Kalinin2019diss} to control the coupling between nearest neighbours in arrays of polaritons condensates. Furthermore, it would be desirable to introduce local dissipation in a reversible manner that can enable both ad-hoc landscape engineering of polariton condensates and real-time control over the coupling between nearest neighbour condensates without adding gain to the system. Despite the potential of such an approach, to date, control of local polariton dissipation has remained illusive both in organic and inorganic microcavities. 

\begin{figure}[t!]
    \includegraphics[scale=1.48]{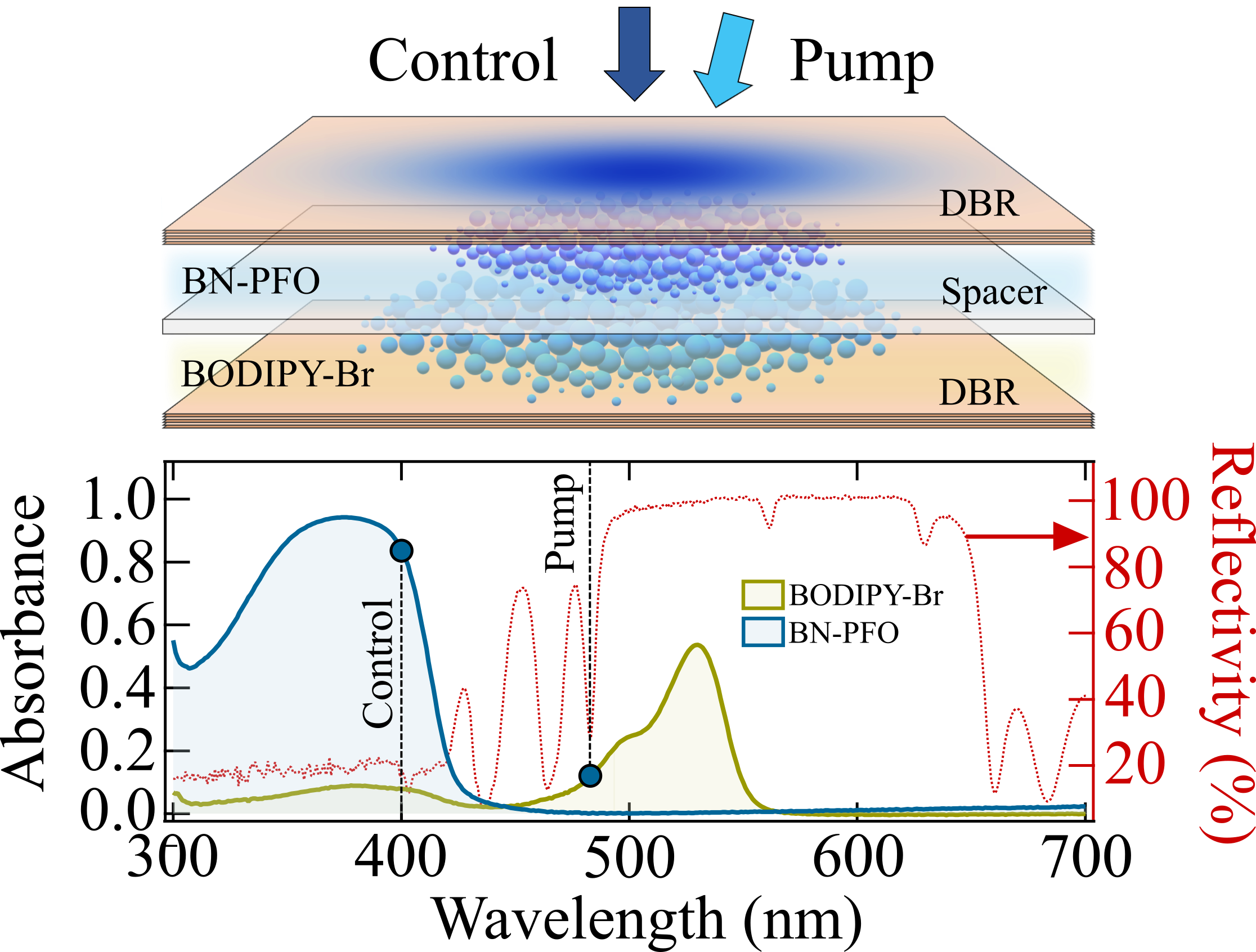}
    \vspace*{-0.25cm} 
    \caption{\label{fig:1} 
        Top: schematic of a double-dye organic microcavity. Bottom panel shows the absorbance spectra of the uncoupled BN-PFO dye molecules and the coupled BODIPY-Br dye molecules (left-axis) and the reflectivity spectrum of the structure (right-axis). Black vertical dashed lines indicate the spectral tuning of pump and control excitation beams.}
\vspace*{-0.5cm}       
\end{figure}

In this Letter, we experimentally demonstrate ultra-fast and reversible manipulation of the condensate energy and density by shaping both the local potential and dissipation in the real space at room temperature. We utilize a double-dye organic microcavity, depicted on the top panel of Fig.\ref{fig:1}, that recently was proposed as a new platform for optically controlled polariton condensate lattices operational at room temperature \cite{Kirsty23arx}. The microcavity includes one layer of strongly coupled BODIPY-Br dye molecules and a blue-detuned uncoupled layer of the copolymer BN-PFO. We demonstrate polariton condensation under non-resonant optical excitation and, by simultaneous excitation of the uncoupled layer, tuning of the condensate energy up to 8 meV. Excited-state absorption of the uncoupled molecular dyes results in increasing losses for polaritons. In tandem with the energy blueshifts, due to the decrease of the intracavity effective refractive index at the energy of the lower polariton mode, excited-state absorption allow us to control the spatial profile of a polariton condensate. 

Polariton density dependent energy blueshifts in semiconductor microcavities are omnipresent at high excitations and usually corroborate evidence of strong coupling. However, the mechanism behind blueshifts in crystalline, usually inorganic, and non-crystalline, usually organic, semiconductors is different. In the first case, the blueshift is predominantly due to the Coulombic exchange interaction between Wannier-Mott excitons \cite{Vladimirova2010, Sun2017}, whereas in the latter case of Frenkel excitons, it is due to the saturation of molecular optical transitions. It was recently shown that the blueshifts in molecular-dye microcavities are not only due to the quenching of the Rabi splitting but predominantly due to changes of the intracavity refractive index from the saturation of both coupled and uncoupled molecules \cite{Yagafarov2020}. Here, we utilise a structure design that allows us to decouple these mechanisms. 

The structure consists of two layers of different organic materials embedded within the microcavity. One 285 nm thick-layer of the molecular dye BODIPY-Br and one 170 nm thick-layer of the conjugated polymer BN-PFO, separated by an inert polyvinyl alcohol (PVA) spacer layer of 60 nm thickness. The organic layers are  sandwiched in between a bottom SiO$_2$/Nb$_2$O$_5$ and a top SiO$_2$/TiO$_2$ distributed Bragg reflector (DBR), as shown schematically in the top panel of Fig.\ref{fig:1}, resulting in quality factor of $\approx 920$; see K. McGhee \emph{et al.} (2023) \cite{Kirsty23arx} for a complete description and characterization of the microcavity structure. The absorption spectra of the BODIPY-Br and BN-PFO layers, and the reflectivity spectrum of the full cavity are shown on the bottom panel of Fig.\ref{fig:1}. Despite the presence of the weakly coupled, blue-detuned to the cavity mode, BN-PFO absorber in the microcavity, the BODIPY-Br layer couples strongly to the bare cavity mode exhibiting a Rabi-splitting ($\Omega$) of 103 meV \cite{Kirsty23arx}. The cavity exhibits a small range of available detunings ($\delta$), and throughout this study the detuning was kept constant at $\approx-$90 meV, with exciton and photon fractions, $|X_{e,p}|^2=(1\pm\delta/\sqrt{(\delta^2+\Omega^2 )})/2$, of $0.17$ and $0.83$, respectively. These two layers are independently excited using a two colour optical excitation beam configuration; see dashed black lines in the bottom panel of Fig.\ref{fig:1}.    

\begin{figure}[t!]
    \includegraphics[scale=1.47]{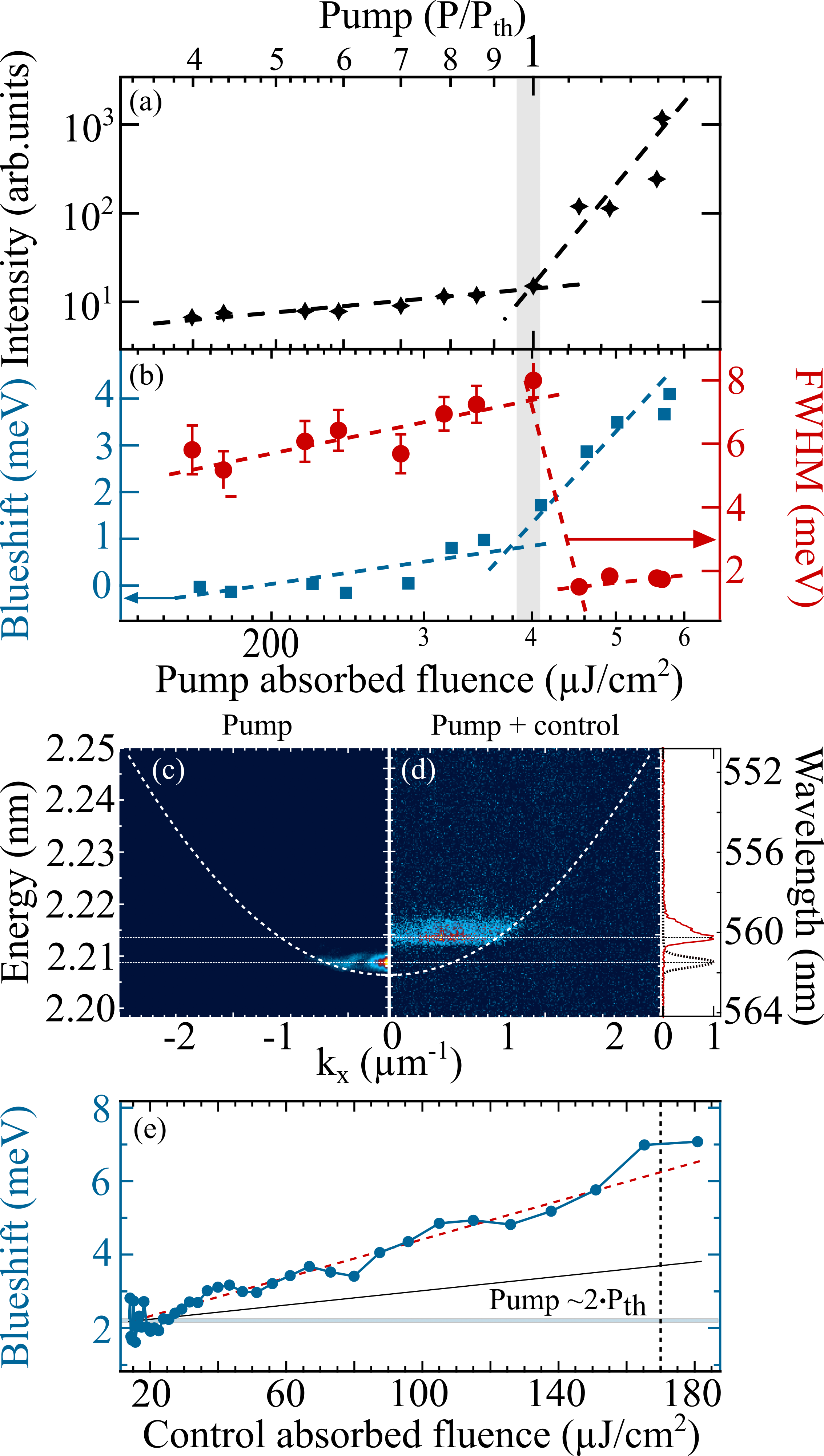}
    \vspace*{-0.25cm} 
    \caption{\label{fig:2} 
        Polariton photoluminescence (PL) \textbf{(a)} intensity, \textbf{(b)} blueshift of the PL spectrum in blue squares (left-axis) and the corresponding linewidth in red dots (right-axis) measured at full width at half maximum (FWHM) vs pump absorbed fluence (bottom-axis) and pump threshold (top-axis). The vertical gray shaded line indicates the condensation threshold. Normalized single-shot dispersion images recorded at $\sim2\times P_{th}$ in the \textbf{(c)} absence and \textbf{(d)} presence of the control beam. The white dashed lines indicate the linear dispersion. The right panel shows the corresponding spectra in black dotted and red solid respectively. \textbf{(e)} Polariton blueshift vs absorbed fluence of the control beam. The horizontal solid-gray line indicates the blueshift of the polariton condensate at $\sim2\times P_{th}$ in the absence of the control beam. The dashed-red line is a linear fit to the blueshift dependence. The black-solid line is the estimated blueshift due to effective refractive index change at the energy of the polariton condensate due to saturation of the molecular optical transitions. The difference in the slope between the black-solid and dashed-red dependencies of the blueshift is attributed to the excited state absorption.}
        \vspace*{-0.5cm}     
\end{figure}

We examine whether polariton condensation is achievable in the double-dye microcavity using non-resonant pulsed optical excitation of 2 ps temporal width tuned at the first Bragg minimum of the reflectivity stop-band; 485 nm. Figure \ref{fig:2}a shows a pump power dependence of polariton photoluminescence (PL) vs the absorbed pump fluence on the bottom horizontal axis. The onset to non-linear emission is defined as the crossing of the two black dashed lines in Fig. \ref{fig:2}a, $P_{th}\approx 400$ $\mu$J/cm$^2$, see bottom horizontal axis. At $P_{th}$, we observe a concomitant collapse of the photoluminescence linewidth, from 8 meV to 1.8 meV, and a blueshift of the emission energy, see Fig.\ref{fig:2}b. Each point on the plots of Figs.\ref{fig:2}a,b is derived from a dispersion image obtained under a single excitation pulse, whose fluence is simultaneously recorded. In Fig.\ref{fig:2}c we show the single-shot dispersion image of the lower polariton branch (LPB) at $P_{pump}\sim2\times P_{th}$. The emission is blueshifted by $2.5$ meV from the linear dispersion that is indicated with a white dotted line. Following the formalism of Ref.\cite{Yagafarov2020}, we can estimate the two contributions to the blueshift from the saturation of the molecular optical transitions for this structure (see also Supplementary Information (SI) Section I). We obtain that the dominant component of the blueshift, 1.84 meV, is due to the effective refractive index renormalisation, and 0.66 meV is due to the vacuum Rabi-splitting quenching. The observed threshold for polariton condensation is similar to that observed previously for BODIPY-family dye filled microcavities \cite{Putintsev2020, Yagafarov2020, Sannikov2019, Cookson2017}, indicating that the presence of the BN-PFO absorber in the cavity does not affect the dynamics of polariton condensation under non-resonant excitation at 485 nm.    

To explore the effect of saturation of the molecular transitions of the second layer of molecular dyes on the polariton condensate, we introduce a second optical excitation ``control'' beam of 500 fs temporal width, and resonant with the absorption of BN-PFO at 400 nm, see the bottom panel of Fig.\ref{fig:1}. We keep the pump fluence of the 485 nm beam constant at twice the condensation threshold, $2\times P_{th}$, and perform a fluence dependence of the control beam at zero time-delay between pump and control pulses. For a control beam fluence of $\approx 250$ $\mu$J/cm$^2$, BN-PFO absorbs $\approx 180$ $\mu$J/cm$^2$, and BODIPY-Br absorbs $\approx 2$ $\mu$J/cm$^2$. The latter corresponds to $\approx 2\%$ of the required absorbed fluence for polariton condensation of BODIPY-Br when pumped at 400 nm \cite{Cookson2017}, which brings a non-discernible effect to the dynamics of the condensate. Under these excitation conditions, we perform single-pulse dispersion imaging, shown in Fig.\ref{fig:2}d, and observe an additional 5 meV blueshift, in the presence of the control pulse. The right panel of Fig.\ref{fig:2}d shows the PL spectra of the condensate in the presence and absence of the control beam. Evidently the linewidth of the condensate in the presence of the control beam is broadened, indicating the onset of dissipation due to excited state absorption from the uncoupled excitons in BN-PFO. Figure \ref{fig:2}e shows the dependence of the blueshift on the absorbed fluence of the control beam. The energy shift dependence on the change of effective refractive index is given by $ \Delta E\cong-E_{c}\frac{\Delta n}{n_{eff}}$, where $\Delta E$ is the energy shift, $E_{c}$ is the bare cavity optical resonance, and $\Delta n$ is the change of the effective cavity refractive index $n_{eff}$ \cite{Yagafarov2020}. For the density of BN-PFO molecules $n_0\approx 1.2\pm0.2$ g/cm$^{-3}$, we can estimate the negative change of the intracavity effective refractive index at the energy of the polariton condensate using the Kramers-Kronig relation and plot the corresponding energy shift with a solid black line in Fig.\ref{fig:2}e, see SI Section I. The difference between the estimated blueshift and the experimental observation is attributed to excited state absorption from the uncoupled molecules that was not previously considered as a contributor to the mechanisms of blueshifts in organic microcavities \cite{Yagafarov2020}. 

\begin{figure}[t!]
    \includegraphics[scale=1.47]{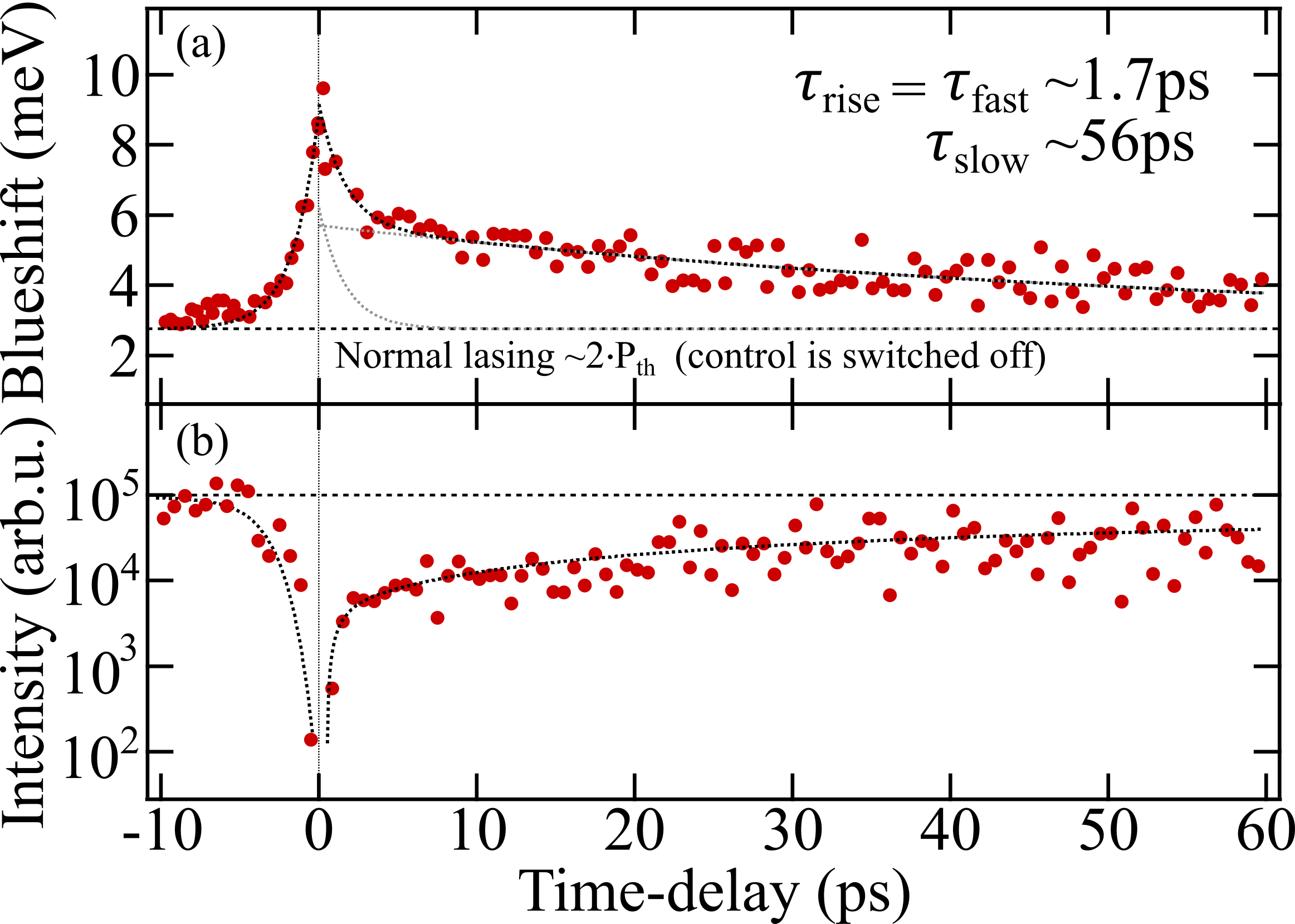}
    \vspace*{-0.5cm} 
    \caption{\label{fig:3} 
        Pump-control time-delay scan: the pump beam fluence is fixed at $\sim2\times P_{th}$ and the control beam absorbed fluence at $\approx 170$ $\mu$J/cm$^2$; positive time-delay corresponds to earlier arrival of the control pulse. \textbf{(a)} Blueshift dependence on the time-delay, exhibiting a double exponential decay with two components shown by gray dashed lines. \textbf{(b)} Polariton PL intensity dependence on the time-delay, exhibiting quenching of emission with the same characteristic times of the blueshift dependence. The horizontal-dashed lines indicate the polariton blueshift, \textbf{(a)}, and intensity, \textbf{(b)}, at $\sim 2\times P_{th}$ in the absence of the control beam; the vertical lines indicate the zero-time delay.}
        \vspace*{-0.5cm} 
\end{figure}

We investigate the transient dynamics of the observed blueshifts by scanning time-delay between the pump and control pulses and recording the dispersion image of the PL for each excitation pulse. We keep the pump fluence constant at $2\times P_{th}$, and the absorbed control beam fluence at $\approx 170$ $\mu$J/cm$^2$, see vertical dashed line in  Fig.\ref{fig:2}e. 
Figure \ref{fig:3}a shows the temporal dependence of the blueshift vs pump-control beam time-delay. The transient decay of the blueshift can be fitted with a bi-exponential decay. The fast component has the same characteristic time, $\approx 1.7$ ps, with the rise dynamics and is dominated by exciton-exciton annihilation in the BN-PFO, resolution limited here by the temporal width of the pump pulses \cite{Marciniak2012}. The slow component, $\approx 56$ ps, corresponds to the recombination dynamics of the excitons in BN-PFO in the absence of non-linear recombination mechanisms. Figure \ref{fig:3}b shows the corresponding temporal dynamics of the condensate PL intensity. Here, we observe strong quenching of the polariton density at zero time-delay that follows approximately the same characteristic times of the blueshift. The depletion of polariton density indicates the onset of strong polariton dissipation in the presence of a far blue detuned and spatially separated exciton reservoir. In the absence of intermolecular energy transfer between the two molecular dyes, the dissipation of polaritons is attributed to excited state absorption of the BN-PFO excitons, as it was also recently revealed through transient absorption spectroscopy for BN-PFO \cite{Kirsty23arx}. 

\begin{figure*}[t!]
    \includegraphics[scale=1.49]{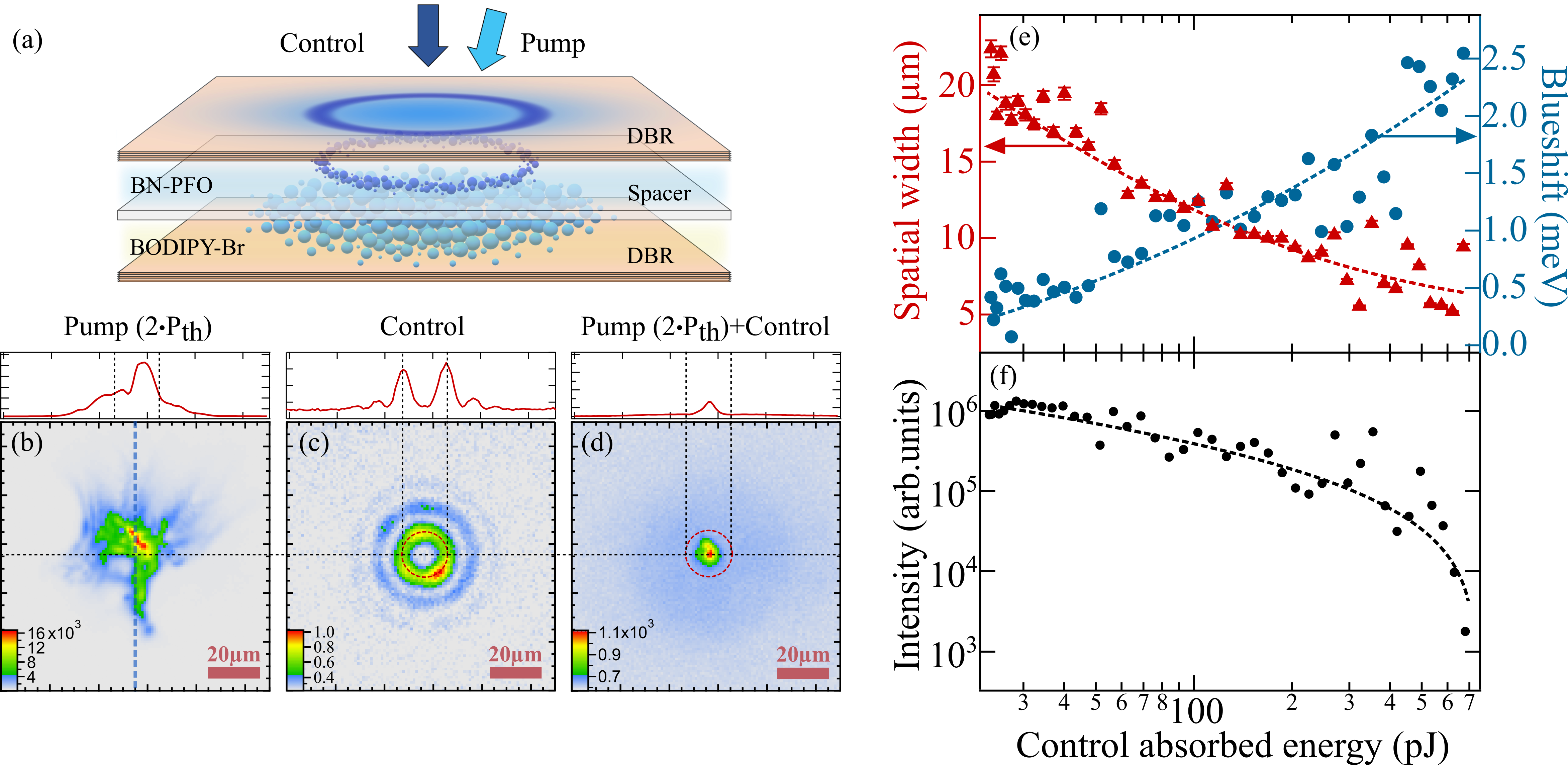}
        \vspace*{-0.25cm}
    \caption{\label{fig:4}   
    \textbf{(a)} Schematic of the Gaussian pump and ring-shaped control excitation beams. The control beam saturates the uncoupled BN-PFO layer at 400nm, and the pump beam non-resonantly injects a polariton condensate in the strongly coupled BODIPY-Br layer at 485nm. \textbf{(b)} Single-shot real-space emission image of a polariton condensation at $2\times P_{th}$. \textbf{(c)} Averaged over 100 realizations real-space emission image of the PL from BODIPY-Br layer excited by $\sim 500$ pJ absorbed energy of the ring-shaped control beam. \textbf{(d)} Single-shot real-space emission image of the polariton condensate of \textbf{(b)} in the presence of the ring-shaped control pulse (red-dashed circle) of \textbf{(c)} at zero time-delay. \textbf{(e)} The dependence of the spatial width (red-markers, left-axis) of the polariton condensate emission across the vertical blue dashed line of \textbf{(b)} and the corresponding blueshift (blue-markers, right-axis) from the centre of the ring of \textbf{(d)} vs the control pulse absorbed energy.  \textbf{(f)} The polariton emission intensity dependence from the center of the ring of \textbf{(d)} vs  the control pulse absorbed energy.}
    \vspace*{-0.5cm} 
\end{figure*}

It is apparent that beyond the control over locally induced blueshifts, excited state absorption induces strong local dissipation, that is both tuneable and reversible. We utilise spatial modulation of the control beam to demonstrate ad-hoc localisation of the polariton condensate. We expand the pump to a $\sim 30$ $\mu$m Gaussian beam at $\approx 2\times P_{th}$ above the condensation threshold, and use a spatial light modulator to shape the control beam into a ring of $\sim20$ $\mu$m in diameter, as shown schematically in Fig.\ref{fig:4}a. Figure \ref{fig:4}b shows the disorder-limited spatial profile of the polariton condensate in the absence of the control beam; the top panel shows the PL intensity cross-section obtained from the horizontal dotted line in Fig.\ref{fig:4}b. Figure \ref{fig:4}c shows the polariton PL in the presence only of the control beam with an absorbed energy of $\sim500$ pJ due to the residual absorption of BODIPY-Br at 400 nm and re-absorption of BN-PFO emission by BODIPY-Br \cite{Kirsty23arx}, and the top panel shows the corresponding horizontal PL intensity cross-section. Figure \ref{fig:4}d shows the respective polariton condensate PL in the presence of both pump and control beams. By tuning the absorbed fluence of the control beam we can simultaneously control both the extent of the condensate and its respective blueshift. Figures \ref{fig:4}e,f show the spatial width, blueshift, and condensate PL intensity vs the control absorbed energy, obtained from a single-pulse spectrally resolved imaging of the condensate real space PL across a vertical cross-section annotated with a blue dashed line in Fig.\ref{fig:4}b. Across the full range of absorbed control fluences we observe that the locally induced dissipation out-competes any pump or control beam induced gain. 

In conclusion, we investigate active spatial control of polariton blueshifts and dissipation by introducing an extra intracavity semiconductor layer, non-resonant to the cavity mode and demonstrate control over the spatial profile of a polariton condensate at room temperature. Our study realises previously unexplored physical mechanisms for the controlled preparation and manipulation of the spatial profile in organic microcavities. 

\begin{acknowledgments}
\vspace*{-0.5cm} 
This work was supported by the Russian Science Foundation (RSF) grant no. 20-72-10145. K.E.M. and D.G.L. thank the U.K. EPSRC for support via the Programme Grant ‘Hybrid Polaritonics’ (EP/M025330/1). K.E.M. thanks the EPSRC for the award of a Doctoral Training Account PhD studentship (EP/R513313/1).
\end{acknowledgments}

\providecommand{\noopsort}[1]{}\providecommand{\singleletter}[1]{#1}%

\setcounter{equation}{0}
\setcounter{figure}{0}
\setcounter{section}{0}
\renewcommand{\theequation}{S\arabic{equation}}
\renewcommand{\thefigure}{S\arabic{figure}}
\onecolumngrid
\newpage
\vspace{1cm}
\begin{center}
\Large \textbf{Supplementary Information}
\end{center}

\section{Section I}

The total polariton blueshift in our double-layered microcavity can be decoupled into for independent components (for details see Ref.\cite{Yagafarov2020}):
\begin{align}\label{eq:1}
 \Delta E_{LPB}&=\Delta E^{\Omega}_{LPB}+\sum_{i}\Delta E^{c_i}_{LPB}+\Delta E^{ESA}_{LPB}= \notag \\
 & =\frac{\xi_1}{2}\frac{s\cdot \hbar\Omega}{\sqrt{1+s^2}}+\sum_{i=1}^{2}\frac{\xi_i}{2}\frac{E_{X}-|\delta|}{5n_{eff}}F[d_i] \cdot\alpha_i\cdot\biggl(1+\frac{1}{\sqrt{1+s^2}}\biggr)+\Delta E^{ESA}_{LPB}
\end{align}

where $\Delta E_{LPB}$ is the total energy shift of LPB that is the sum of energy shifts due to vacuum Rabi-splitting quenching in strongly coupled layer, $\Delta E^{\Omega}_{LPB}$, cavity mode renormalization with contributions to the refractive index modulation from both organic layers, with $i=1,2$ relating to BODIPY-Br and BN-PFO, respectively, and $\Delta E^{ESA}_{LPB}$ is the energy shift caused by refractive index modulation due to the excited state absorption effect in BN-PFO , $\xi$ is the saturation parameter, $s=\hbar\Omega/|\delta|$ is a dimensionless parameter of strong coupling, $\alpha$ corresponds to the oscillator strength of the optical transition, $E_{X}$ is the BODIPY-Br exciton resonance, $\delta=E_c-E_x$ is the detuning, and $F[d]$ is the Dawson function (integral), $F[d]=e^{-d^2}\int_{0}^{d}e^{x^2}dx$. 

Lets consider two organic layers separately. In strongly coupled BODIPY-Br layer in the absence of the control beam the emission is blueshifted by $2.5$ meV from the linear dispersion. Following the formalism of Ref.\cite{Yagafarov2020}, we can estimate the two contributions to the blueshift from the saturation of the molecular optical transitions for this structure. Taking the ratio of two first terms in the Eq.\ref{eq:1} we obtain independent of $\xi$ expression:

\begin{equation}\label{eq:2}
    \rho =\frac{\Delta E^{c_1}_{LPB}}{\Delta E^{\Omega}_{LPB}}=\frac{(E_{X}-|\delta|)\cdot F[d_1]\cdot\alpha_1\cdot(\sqrt{s^2+1}+1)}{5n_{ff}\cdot s\cdot\hbar\Omega_{0}}  
\end{equation}

Here $E_{X}=2.34672$ eV $(528.33$ nm) is the BODIPY-Br main exciton line, $\hbar\Omega_{0}=103$ meV is the Rabi-splitting, $|\delta|=|549.18-528.33|=89.1$ meV is the detuning, 
$\alpha_1=Abs_{max}\cdot\lambda_{max}/L_{layer}=0.55182\cdot528.33/285$ [$1\cdot$nm/nm] is the oscillator strength of the optical transition, $s=\hbar\Omega_{0}/|\delta|=103/89.1$ [meV/meV] is a dimensionless parameter of strong coupling, $d_1=2\sqrt{ln(2)}\cdot\delta/FWHM=2\sqrt{ln(2)}\cdot89.1/127.181=1.16654$, and $F[1.16654]=0.514261$, $n_{eff}=1.81$. Substituting these values into \ref{eq:2}, we get:

\begin{equation}\label{eq:3}
    \rho= 2.78684
\end{equation}

It allows us to obtain components of the blueshift in BODIPY-Br layer, in the absence of the control beam, due to the effective refractive index renormalisation, which is 1.84 meV, and due to the vacuum Rabi-splitting quenching, which is 0.66 meV.

The uncoupled BN-PFO layer contributes to the blueshift through effective refractive index change due to two effects: the saturation of the molecular optical transitions and excited state absorption on top of those saturated transitions. The first effect is again described by the second term of Eq.\ref{eq:1} and scales linearly with the saturation parameter $\xi_2$. To estimate its absolute contribution to the additional blueshift we have to use parameters relating to BN-PFO layer: $\alpha_2=Abs_{max}\cdot\lambda_{max}/L_{layer}=0.94\cdot371/170$ [$1\cdot$nm/nm] is the oscillator strength of the optical transition, $d_2=2\sqrt{ln(2)}\cdot\delta/FWHM=2\sqrt{ln(2)}\cdot1084.3/858=2.10429$ and $F[2.10429]=0.281093$, $n_{eff}=1.81$.
Substituting these values into the following expression:

\begin{equation}\label{eq:4}
    \Delta E^{c_2}_{LPB}=\frac{\xi_2}{2}\frac{E_{X}-|\delta|}{5n_{eff}}F[d_2] \cdot\alpha_2\cdot\biggl(1+\frac{1}{\sqrt{1+s^2}}\biggr)  
\end{equation}

We obtain:

\begin{equation}\label{eq:5}
    \Delta E^{c_2}_{LPB}\cong112 [meV]\cdot\xi_2  
\end{equation}

\begin{figure}[t]
    \includegraphics[scale=2]{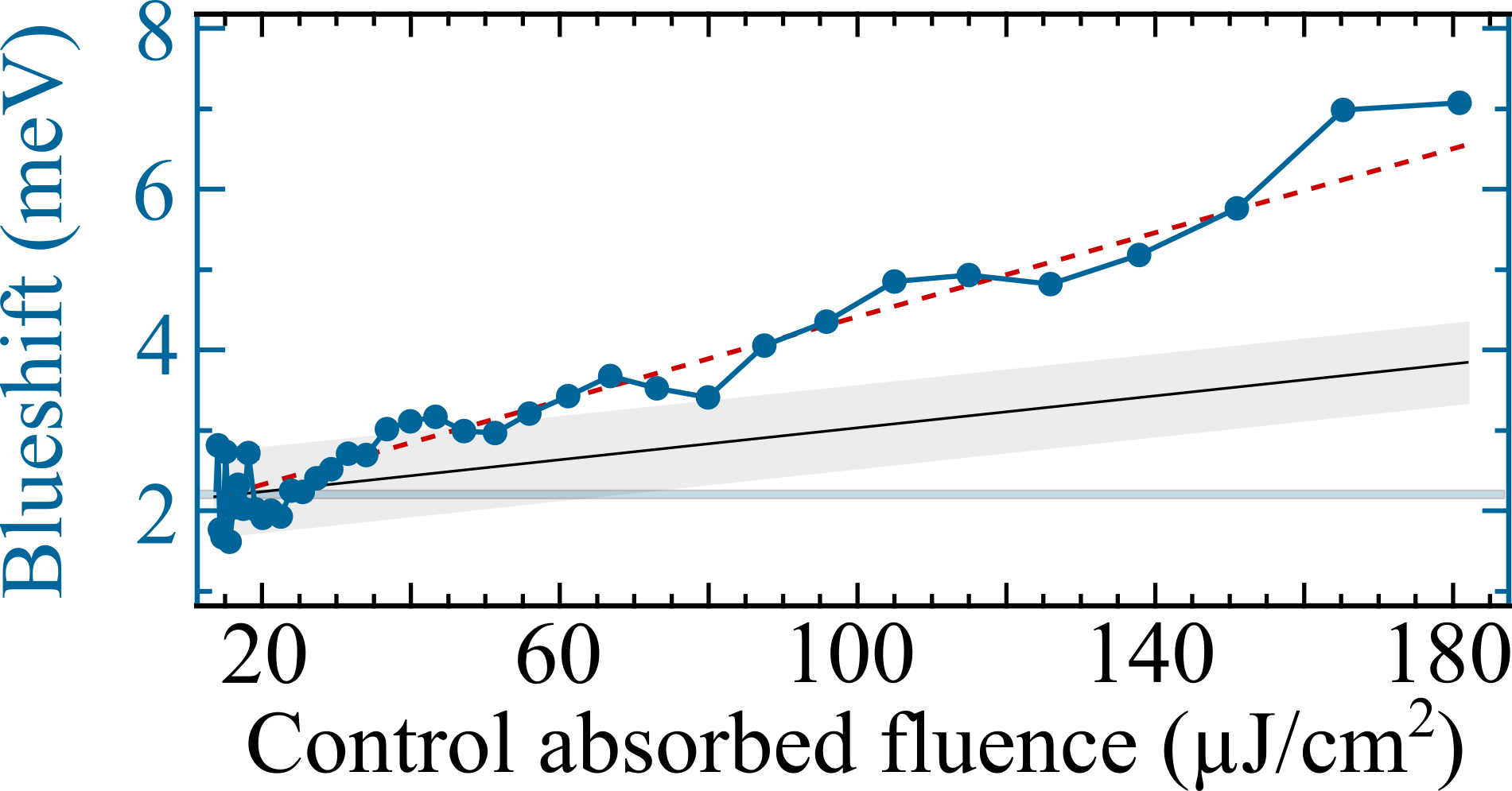}
    \centering
    \caption{\label{SI:1} 
         Polariton blueshift vs absorbed fluence of the control beam. The horizontal solid-gray line indicates the blueshift of the polariton condensate at $\sim2\times P_{th}$ in the absence of the control beam. The dashed-red line is a linear fit to the blueshift dependence. The black-solid line is the estimated blueshift due to effective refractive index change at the energy of the polariton condensate due to saturation of the molecular optical transitions with the error gray-shaded error bar. The difference in the slope between the black-solid and dashed-red dependencies of the blueshift is attributed to the excited state absorption.}
\end{figure}

This expression defines the dependence of the estimated blueshift due to effective refractive index change at the energy of the polariton condensate on the saturation of the molecular optical transitions $\xi_2$. With the control pulse absorbed fluence in the BN-PFO layer at 400 nm of $\approx 200$ $\mu$J/cm$^2$, the number of excited molecules is $N\sim 10^9$ within a volume of 53 $\mu$m$^{3}$ (20 $\mu m$ spot-size and 170nm-thick layer). The estimated molecular density in BN-PFO layer is $n_0\approx 1.2\pm0.2$ g/cm$^{-3}$, which, given the weight of a polymer unit of M$_n=$68300 \cite{Kirsty23arx}, results in $N_0\approx 5.6\cdot10^{8}$ available polymer chains within the same 53 $\mu$m$^{3}$ volume-size. However, the number of chromophores is assumed to be higher \cite{Schindler2005}, since the chromophores are localized to shorter conjugated segments (the so-called “effective conjugation length”), and this length is different for individual polymers. One chain carry multiple chromophores. For BN-PFO, the binaphthyl units that are randomly incorporated into the chains are expected to separate oligo/polyfluorene segments (of different length), and should act as conjugation barriers thus separating individual PFO chromophores. Thus the number of chromophores can be higher by a factor of 10 (or more). With this in mind, we estimate the number of available absorbing units $N_0\approx 5.6\cdot10^{10}$ resulting in the saturation parameter $\xi_2\approx N/N_0=10^9/5.6\cdot10^{10}\approx 0.018$ at $200$ $\mu$J/cm$^2$ control absorbed fluence. Fig.S1 shows the resulting dependence of the blueshift due to the saturation of the molecular optical transitions on the control absorbed fluence, black-solid line. The difference in the slope between the black-solid and dashed-red dependencies of the blueshift is $\Delta E^{ESA}_{LPB}$ which is attributed to excited state absorption.

\end{document}